\documentclass[runningheads]{llncs}

\usepackage{amssymb}
\usepackage{amsmath}
\usepackage{url}
\usepackage{graphics}
\usepackage{graphicx}
\usepackage[all]{xy}
\usepackage{float}

\usepackage{algorithmic}
\usepackage[ruled,vlined]{algorithm2e}
\usepackage{alltt}
\urldef{\mailsa}\path|pat@dcs.gla.ac.uk|

\newlength{\halftextwidth}
\usepackage{multicol}
\usepackage{listings}
\usepackage{amsopn}

\begin{document}

\title{Empirical Algorithmics: draw your own conclusions}
\titlerunning{Draw your own conclusions}
\toctitle{Empirical Algorithmics: draw your own conclusions}
\author{F. Prefect and P. Prosser}
\authorrunning{Prefect and Prosser}
\tocauthor{Ford Prefect and Patrick Prosser}
\institute{School of Computing Science, University of Glasgow}
\maketitle

\begin{abstract}
In an empirical comparisons of algorithms we might compare run times over a set of benchmark
problems to decide which one is fastest, i.e. an algorithmic horse race. Ideally we would like to
download source code for the algorithms, compile and then run on our machine. Sometimes code isn't
available to download and sometimes resource isn't available to implement all the algorithms we want
to study. To get round this, published results are rescaled, a technique endorsed by DIMACS, and those rescaled results included in a new study. 
This technique is frequently used when presenting new algorithms for the maximum clique problem.
We demonstrate that this
 is unsafe, and that if carelessly used may allow us
to draw conflicting conclusions from our empirical study.
\end{abstract}

\section{Introduction}
How can we determine if one algorithm ($A$) is better than other ($B$)? 
Most often we will put them to the test in the guise of a simple beauty contest, or what
David Johnson has called a ``horse race'' \cite{howTo}. Given a set of benchmarks apply algorithms $A$ and $B$ to problems and measure the 
number of times the \emph{principal activity} of the algorithm is executed and the run time, on each problem instance. 
But how can we get accurate and reliable measures of run time performance?

Run times of an algorithm are influenced by many factors such as programming language, compiler, machine, operating system
and of course programmer ability! We would like efficient implementations of algorithms $A$ and $B$ to run on our machine 
against a recognised set of publicly available benchmark problems. Ideally we would like to download source code for $A$ and $B$
and compile and run on our machine. 
But often this is not possible because a published result used 
a commercial toolkit (such as in \cite{regin2003}) that we cannot afford or the policy adopted by an employer or laboratory
prevents the release of code. Therefore we might implement both algorithms ourselves, and then
make our code and results available to all. 
But what if we do not have the time or the ability to implement an algorithm in our study?
What we might do is compare against published results, taking into consideration speed differentials between our machine and that used in the 
publication, rescale results accordingly and then use those results in our study. This is an approach proposed in the Second DIMACS Implementation Challenge \cite{dimacs93}.


In this paper we examine two state of the art algorithms for the maximum clique problem. They are applied to a recognised 
set of benchmarks and we conclude that one algorithm is faster than the other. We then pretend that only
one of these implementations is available to us and then use the rescaling technique that is widely used \cite{dimacs93}, i.e.
we run two different algorithms on two different machines and scale run times based on a benchmark program and problems, i.e. we put the rescaling technique
to the test. We use three different machines and two programming languages and show that rescaling produces wildly inaccurate results and worse still,
allows us to show that our fastest algorithm is the slowest: we demonstrate that using this
rescaling technique we can draw whatever conclusion we want.

\section{The Maximum Clique Problem and Two Algorithms}
\vspace{-1.5mm}
We now present the maximum clique problem and give a brief overview of exact algorithms for this problem. We then present 
the two algorithms used in this study.

\subsection{The maximum clique problem}
A \emph{graph} $G$ consists of a pair of sets $(V,E)$, where $V$ is the set of \emph{vertices} and $E \subseteq V \times V$ is the set of \emph{edges}. We say vertices $u$ and $v$ are \emph{adjacent} if $\{u, v\} \in E$. Throughout, our graphs are \emph{simple} (no vertex is adjacent to itself), and \emph{undirected} (if $u$ is adjacent to $v$ then $v$ is adjacent to $u$). A \emph{clique} is a set of vertices $C \subseteq V$ such that every distinct pair of vertices in $C$ is adjacent in $G$.
Finding a clique in a given graph is one of the six basic NP-complete problems given in \cite{gareyJohnson}. It is posed as a decision problem [GT19]: given a simple undirected graph $G = (V,E)$ and a positive integer $k \leq |V|$, does $G$ contain a clique of size $k$ or more? The optimization problems is then to find a \emph{maximum clique}, whose size we denote $\omega(G)$.

A graph can be coloured, by that we mean that any pair of adjacent vertices must be given different colours. We do not use colours, 
we use integers to label the vertices. The minimum number of different colours required is then the
\emph{chromatic number} of the graph $\chi(G)$, and $\omega(G) \leq \chi(G)$. Finding the chromatic number is NP-complete.

We can address the decision and optimization problems with an exact algorithm, 
such as a backtracking search 
\cite{fahle,regin2003,wood97,carraghanPardalos90,pardalosRodgers92,prjo2002,segundo2011,segundo2011b,Konc_Janezic_2007,tomita2003,tomita2010,aaai2010,carmoZuge}.
Backtracking search incrementally constructs the set $C$ (initially empty) by choosing a \emph{candidate vertex}
from  the \emph{candidate set} $P$ (initially all of the vertices in $V$) and then adding it to $C$. 
Having chosen a vertex the candidate set is then updated, removing vertices
that cannot participate in the evolving clique. If the candidate set is empty then $C$ is maximal (and if it is a maximum we save it)
and we then backtrack. Otherwise $P$ is not empty and we continue our search, selecting from $P$ and adding to $C$.
There are other scenarios where we can cut off search, e.g. if what is in $P$ is insufficient
to unseat the champion (the largest clique found so far) search can be abandoned, i.e. an upper bound can be computed.
Graph colouring can be used to compute an upper
bound during search; if the candidate set can be coloured with $k$ colours then it can contain a clique no larger than $k$
\cite{wood97,fahle,segundo2011,Konc_Janezic_2007,tomita2003,tomita2010}. 

\subsection{MCSa and BBMC}
In our study we compare two exact algorithms. The first is MCSa  and
corresponds to MCSa1 in \cite{exactProsser} and is essentially Tomita's MCS \cite{tomita2010} with the colour 
repair step removed. Our second algorithm is BB-MaxClique \cite{segundo2011} and we will call it BBMC (where BB abbreviates ``Bit Board'').
It is at heart a bit-set encoding of  MCSa with the design goal of exploiting bit parallelism, and is again reported
in \cite{exactProsser}. We describe MCSa and BBMC with the aid of the \emph{generic} algorithm MC (Algorithm \ref{mc}).

Algorithm $MC$ explores a backtrack tree to find a largest clique in $G$ via a call (line 5) to procedure $expand$ (lines 7 to 18) 
and takes three arguments:
the growing clique $C$ (initially empty), 
the candidate set $P$ (initially all the vertices in the graph sorted in non-increasing degree order,
to improve sequential colouring) and 
the graph $G$.
Search starts by sequentially colouring the graph \cite{welshPowell} induced by $P$ (line 9), delivering a pair $(S,colour)$ where $S$ is a stack
of vertices, $colour$ is an array of colours and the vertices in the stack are in colour order (highest coloured vertex at top of stack).
If Vertex $v$ is at the top of the stack then vertex $v$ has colour $colour_{v}$ and 
all vertices in the stack have a colour less than or equal to $colour_{v}$. Consequently the graph induced by the
vertices in $S$ can be coloured with $colour_{v}$ colours and must then have a clique no larger than $colour_{v}$.
The vertex $v$ is popped from $S$ (line 11) and if the colour of that vertex is too small then
the graph induced by $v$ and the remaining vertices in $S$ and the vertices in the growing clique $C$
will be insufficient to unseat the current champion and search can be terminated (line 12). 
Otherwise the vertex $v$ is added to the clique $C$ (line 13) 
and a new candidate set is produced $P'$  (line 14) where $P'$ is the set of vertices in $P$ that are adjacent to the current vertex $v$
(where $N(v,G)$ delivers the set of vertices adjacent to $v$ in the graph $G$, i.e. the $N$eighbourhood of $v$).
Consequently each vertex in $P'$ is adjacent to all vertices in $C$. If the new candidate set is empty then $C$ is maximal
and if it is larger than the largest clique found so far it is saved (line 15). But if $P'$ is not empty $C$ is not maximal
and $C$ can grow via recursive calls to $expand$. Regardless, when all possibilities of expanding the 
current clique with the vertex $v$ have been considered that vertex can be removed from the current clique
and from the candidate set (lines 17 and 18).

\begin{algorithm}
\DontPrintSemicolon
\nl $\textbf{Set} ~ MC(\textbf{Graph} ~ G)$ \;
\nl \Begin{
\nl $\textbf{Global} ~ n ~ \gets |V(G)|$ \;
\nl $\textbf{Global} ~ C_{max} ~ \gets \emptyset$ \;
\nl $expand(\emptyset,sort(V(G),G),G)$ \;
\nl $\textbf{return} ~ C_{max}$ \;
}
\;
\nl $\textbf{void}~expand(\textbf{Set}~C,\textbf{Set}~P,\textbf{Graph}~G)$ \;
\nl \Begin{
\nl $(S,colour) \gets colourSort(P,G)$ \;
\nl \While {$S \neq \emptyset$}{
\nl $v \gets pop(S)$ \;
\nl \lIf {$colour_{v} + |C| \leq |C_{max}|$}{$\textbf{return}$} \;
\nl $C \gets C \cup \{v\}$ \;
\nl $P' \gets P \cap N(v,G)$ \;
\nl \lIf {$P' = \emptyset$~\bf{and}~$|C| > |C_{max}|$}{$C_{max} \gets C$} \;
\nl \lIf {$P' \neq \emptyset$}{$expand(C,P',G)$} \;
\nl $C \gets C \setminus \{v\}$ \;
\nl $P \gets P \setminus \{v\}$ \;
 }
}
\caption{The generic maximum clique algorithm MC}
\label{mc}
\end{algorithm}


In BBMC all sets are represented as bit strings. Initially the graph is permuted, i.e. vertices are renamed such that they are in 
non-decreasing degree order and this is done to improve sequential colouring (this replaces the $sort$ in line 5). 
Search in BBMC is identical to that in MCSa with the exception of  set union and intersection operations (lines 13 and 14): these are 
replaced with logical bit operations in an attempt to exploit bit-parallelism. There is also a 
difference in the implementation of the vertex colouring algorithm ($colourSort$, line 9): in MCSa colouring iterates over vertices assigning 
a colour to each in turn; in BBMC colour classes are iterated over adding un-coloured non-adjacent vertices to the current colour class
(the colour class is then an independent set).

\section{MCSa versus BBMC: a maximum clique horse race}
\vspace{-1.5mm}
We now report on ``the horse race'' to decide which algorithm is better. We use the DIMACS instances \cite{DIMACS},
the default benchmark suite used for algorithmic comparison of maximum clique algorithms.
The algorithms were coded up in Java \mbox{version 1.6.0\_07} and experiments run on our reference machine (named Cyprus), 
a machine with two Intel E5620 2.4~GHz quad-core processors with 48 GB memory, running Linux CentOS 5.3.

In Table \ref{tableMCSvBBMC} we tabulate \emph{Goldilocks} instances from the benchmark suite: we remove the instances that are too
easy (take less than a second, i.e. Pet Peeve 3 \cite{howTo}) and those that are too hard (take more than 4 hours, Pet Peeve 8 \cite{howTo}) 
leaving those that are ``just right'' for both algorithms. We tabulate the number of calls to $expand$ (and this is the same for 
both algorithms), the size of the maximum clique 
($\omega$), run time in whole seconds for MCSa and BBMC, and on the far right the ratio of MCSa's run time over BBMC's run time
(and a value greater than 1 shows that BBMC was faster by that amount). What we see is what we expect: BBMC is the algorithm of 
choice, typically twice as fast as MCSa\footnote{Obviously, although twice as fast on these benchmarks, this might not hold on
other instances.}.

\begin{table}
\begin{center}
\begin{scriptsize}
\begin{tabular}{|l | r r | r r |c|} \hline 
instance & $expand$ & $\omega$ & MCSa & BBMC & MCSa/BBMC \\ \hline
brock200-1 & 524,723 & 21 & 4 & 2 & 2.03   \\ 
brock400-1 & 198,359,829 & 27 & 2,888 & 1,421 & 2.03   \\ 
brock400-2 & 145,597,994 & 29 & 2,089 & 1,031 & 2.03   \\ 
brock400-3 & 120,230,513 & 31 & 1,616 & 808 & 2.00   \\ 
brock400-4 & 54,440,888 & 33 & 802 & 394 & 2.03   \\ 
brock800-4 & 640,444,536 & 26 & 12,568 & 6,908 & 1.82   \\ 
MANN-a27 & 38,019 & 126 & 6 & 1 & 4.12   \\ 
MANN-a45 & 2,851,572 & 345 & 3,766 & 542 & 6.94   \\ 
p-hat1000-1 & 176,576 & 10 & 2 & 1 & 1.80   \\ 
p-hat1000-2 & 34,473,978 & 46 & 1,401 & 720 & 1.95   \\ 
p-hat1500-1 & 1,184,526 & 12 & 14 & 9 & 1.52   \\ 
p-hat300-3 & 624,947 & 36 & 13 & 5 & 2.36   \\ 
p-hat500-2 & 114,009 & 36 & 3 & 1 & 2.56   \\ 
p-hat500-3 & 39,260,458 & 50 & 1,381 & 606 & 2.28   \\ 
p-hat700-2 & 750,903 & 44 & 27 & 12 & 2.20   \\ 
san1000 & 150,725 & 15 & 10 & 5 & 1.76   \\ 
san200-0.9-2 & 229,567 & 60 & 5 & 2 & 2.36   \\ 
san200-0.9-3 & 6,815,145 & 44 & 111 & 50 & 2.20   \\ 
san400-0.7-1 & 119,356 & 40 & 2 & 1 & 2.04   \\ 
san400-0.7-2 & 889,125 & 30 & 19 & 9 & 2.12   \\ 
san400-0.7-3 & 521,410 & 22 & 10 & 5 & 2.10   \\ 
san400-0.9-1 & 4,536,723 & 100 & 422 & 125 & 3.37   \\ 
sanr200-0.9 & 14,921,850 & 42 & 283 & 123 & 2.30   \\ 
sanr400-0.5 & 320,110 & 13 & 2 & 1 & 1.85   \\ 
sanr400-0.7 & 64,412,015 & 21 & 711 & 365 & 1.95   \\ \hline
\end{tabular}
\end{scriptsize}
\end{center}
\caption{MCSa versus BBMC: DIMACS \emph{Goldilocks} instances, calls to $expand$, size of largest clique, run time in seconds for MCSa and BBMC, 
relative performance of the algorithms. Conclusion: BBMC is twice as fast as MCSa.}
\label{tableMCSvBBMC}
\end{table}

\section{Calibration of results}
\vspace{-1.5mm}
For the maximum clique problem authors compile and run a standard C program, dfmax, against a 
set of benchmarks. These run times are then used as a conversion factor, 
and results are then taken from one publication, scaled accordingly, and then included in another publication. This 
technique has been used for more than a decade. Some recent examples of this are
\cite{prjo2002} including rescaled results from \cite{sewell98};
\cite{regin2003} including rescaled results from \cite{prjo2002}, \cite{wood97} and \cite{fahle};
\cite{tomita2007} including rescaled results from \cite{prjo2002}, \cite{fahle} and \cite{sewell98};
\cite{tomita2010} including rescaled results from \cite{regin2003} and \cite{prjo2002};
\cite{segundo2011} including rescaled  results from \cite{Konc_Janezic_2007}; 
\cite{segundo2011b} including rescaled  results from \cite{segundo2011} and \cite{tomita2010};
\cite{aaai2010} including rescaled results from \cite{tomita2007} and \cite{regin2003} and most recently
\cite{MCMD14} including results from \cite{segundo2011}, \cite{segundo2011b} and \cite{tomita2010}.
These rescalings are presented in Figure 1 as a directed graph where a directed edge $v \longrightarrow w$ means 
\emph{rescaled results from paper [v] are included in paper [w]}.

\begin{figure}
\begin{center}
 \includegraphics[width=0.9\textwidth]{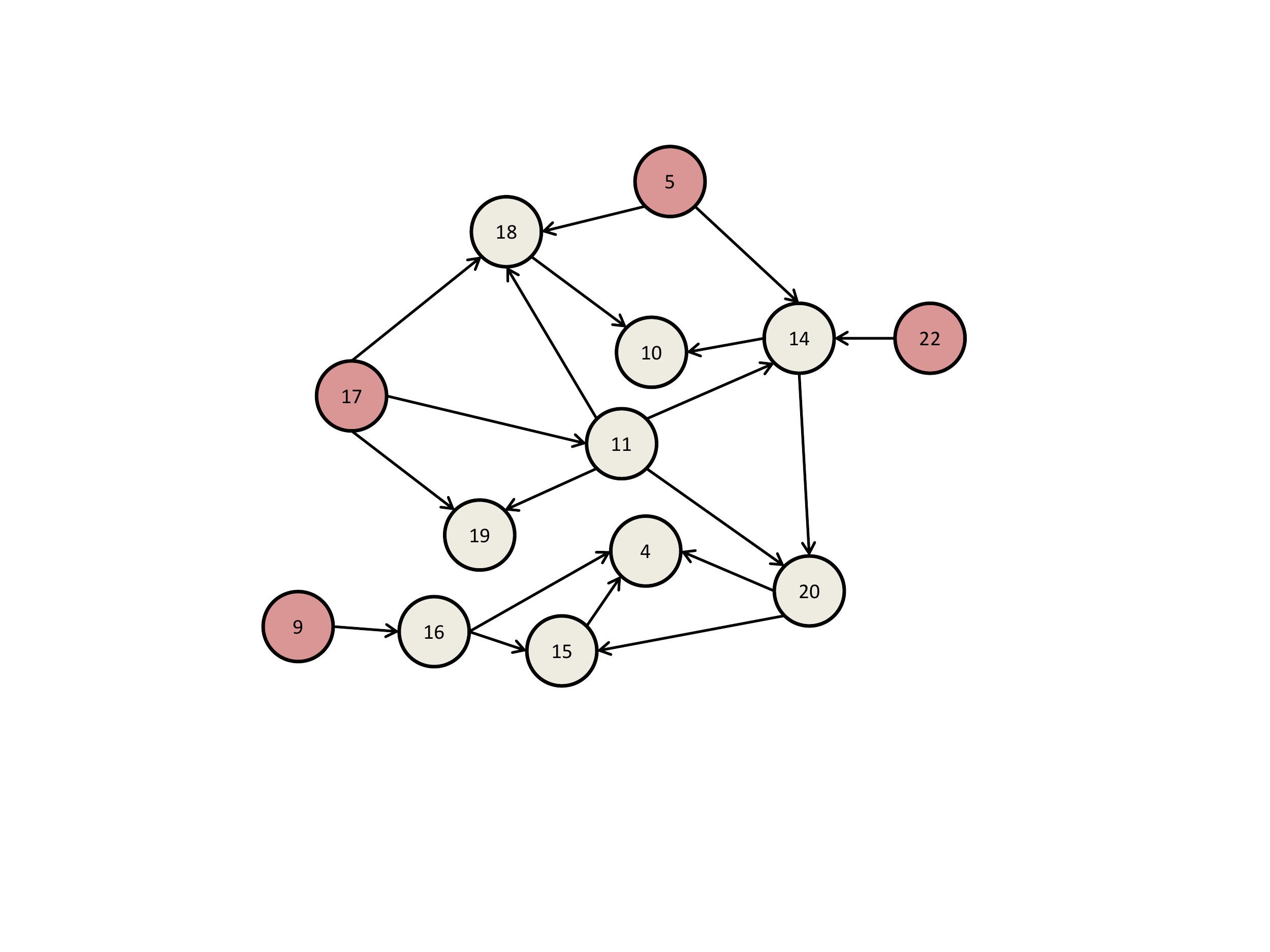}
\vspace{-20mm}
 \caption{Rescalings. A numbered vertex corresponds to a paper in the References. A directed edge $(v,w)$ means that paper [w] includes rescaled results from paper [v]. Vertices with in-degree
zero do not include rescaled results and are coloured pink.}
 \end{center}
\label{graph}
\end{figure}

Is this rescaling procedure safe?
To answer this we take two additional machines, Fais and Daleview, and calibrate them with respect to our reference machine Cyprus. 
We then run experiments 
on each machine using the Java implementations of the algorithms against some of the DIMACS benchmarks. These results are
then rescaled. If the rescaling gives substantially different results from those on the reference machine
this would suggest that this technique is not safe.

Table \ref{rosettaStone} gives a ``Rosetta Stone'' for the three machines used in this study.
The standard program dfmax \footnote{Available from ftp://dimacs.rutgers.edu/pub/dsj/clique} was compiled using 
gcc and the -O2 compiler option on each machine and then run
on the benchmarks r* on each machine. Run times in seconds are tabulated for the five benchmark instances,
each machine's /proc/cpuinfo is given and a conversion factor relative to the reference machine Cyprus
is then computed in the same manner as that reported in
\cite{segundo2011} \emph{``... the first two graphs from the benchmark were removed (user time was considered too small) and the rest
of the times averaged ...''.}  Therefore when rescaling the run times from Fais we
multiply actual run time by 0.41 and for Daleview by 0.50.

\begin{table}
\begin{center}
\begin{scriptsize}
\begin{tabular}{|l|c|c|c|c|c|c|c|c|c|c|} \hline 
machine & r100.5 & r200.5 & r300.5 & r400.5 & r500.5 & Intel(R) & GHz & cache & Java & scaling factor\\ \hline
Cyprus & 0.0 & 0.02 & 0.24 & 1.49 & 5.58 & Xeon(R) E5620 & 2.40 & 12,288KB & 1.6.0\_07 & 1 \\ \hline
Fais & 0.0 & 0.08 & 0.58 & 3.56 & 13.56 & XEON(TM) CPU & 2.40& 512KB & 1.5.0\_06 & 0.41 \\ \hline
Daleview & 0.0 & 0.09 & 0.53 & 3.00 & 10.95 & Atom(TM) N280 & 1.66 & 512KB & 1.6.0\_18 & 0.50 \\ \hline
\end{tabular}
\end{scriptsize}
\end{center}
\caption{Conversion factors using dfmax on three machines: Cyprus, Fais and Daleview}
\label{rosettaStone}
\end{table}

\begin{table}
\begin{center}
\begin{scriptsize}
\begin{tabular}{|l|c c|c c|c c|c c|c c|c c|} \hline 
\multicolumn{1}{|c|} {} & \multicolumn{6}{|c|}{MCSa} & \multicolumn{6}{|c|}{BBMC}\\
\multicolumn{1}{|c|} {instance} & \multicolumn{2}{|c}{Fais} & \multicolumn{2}{c}{Daleview} & \multicolumn{2}{c|}{Cyprus} & \multicolumn{2}{|c}{Fais} & \multicolumn{2}{c}{Daleview} & \multicolumn{2}{c|}{Cyprus} \\ \hline
brock200-1 & 0.25 & (19.3) & 0.27 & (17.5) & 1.00 & (4.8) & 0.15 & (15.4) & 0.09 & (25.0) & 1.00 & (2.4) \\ 
brock200-4 & 0.40 & (1.8) & 0.43 & (1.7) & 1.00 & (0.7) & 0.20 & (1.6) & 0.13 & (2.5) & 1.00 & (0.3) \\ 
hamming10-2 & 0.18 & (1.9) & 0.14 & (2.3) & 1.00 & (0.3) & 0.25 & (0.6) & 0.21 & (0.7) & 1.00 & (0.2) \\ 
hamming8-4 & 0.24 & (1.9) & 0.28 & (1.6) & 1.00 & (0.5) & 0.23 & (1.6) & 0.19 & (1.9) & 1.00 & (0.4) \\ 
johnson16-2-4 & 0.35 & (2.3) & 0.38 & (2.2) & 1.00 & (0.8) & 0.26 & (1.9) & 0.14 & (3.5) & 1.00 & (0.5) \\ 
MANN-a27 & 0.21 & (32.3) & 0.22 & (31.9) & 1.00 & (6.9) & 0.14 & (12.3) & 0.10 & (16.5) & 1.00 & (1.7) \\ 
p-hat1000-1 & 0.25 & (8.4) & 0.28 & (7.4) & 1.00 & (2.1) & 0.14 & (8.4) & 0.12 & (9.4) & 1.00 & (1.2) \\ 
p-hat1500-1 & 0.19 & (77.8) & 0.22 & (66.1) & 1.00 & (14.4) & 0.11 & (90.4) & 0.10 & (92.2) & 1.00 & (9.5) \\ 
p-hat300-3 & 0.25 & (53.4) & 0.26 & (51.0) & 1.00 & (13.5) & 0.14 & (41.7) & 0.09 & (60.1) & 1.00 & (5.7) \\ 
p-hat500-2 & 0.27 & (13.4) & 0.30 & (12.1) & 1.00 & (3.7) & 0.14 & (10.2) & 0.11 & (13.4) & 1.00 & (1.4) \\ 
p-hat700-1 & 0.40 & (1.6) & 0.51 & (1.3) & 1.00 & (0.6) & 0.29 & (1.2) & 0.24 & (1.4) & 1.00 & (0.3) \\ 
san1000 & 0.11 & (94.1) & 0.12 & (89.3) & 1.00 & (10.5) & 0.10 & (57.9) & 0.11 & (54.8) & 1.00 & (5.9) \\ 
san200-0.9-1 & 0.29 & (4.9) & 0.31 & (4.7) & 1.00 & (1.4) & 0.18 & (4.2) & 0.11 & (6.6) & 1.00 & (0.7) \\ 
san200-0.9-2 & 0.22 & (23.5) & 0.25 & (20.9) & 1.00 & (5.2) & 0.15 & (14.6) & 0.09 & (23.6) & 1.00 & (2.2) \\ 
san400-0.7-1 & 0.25 & (10.2) & 0.27 & (9.6) & 1.00 & (2.6) & 0.15 & (8.3) & 0.12 & (10.2) & 1.00 & (1.3) \\ 
san400-0.7-2 & 0.23 & (84.2) & 0.27 & (72.9) & 1.00 & (19.6) & 0.13 & (71.4) & 0.11 & (87.3) & 1.00 & (9.2) \\ 
san400-0.7-3 & 0.24 & (45.5) & 0.27 & (40.8) & 1.00 & (10.8) & 0.13 & (39.8) & 0.11 & (46.8) & 1.00 & (5.2) \\ 
sanr200-0.7 & 0.31 & (5.0) & 0.33 & (4.7) & 1.00 & (1.5) & 0.19 & (4.1) & 0.12 & (6.7) & 1.00 & (0.8) \\ 
sanr200-0.9 & 0.23 & (1,249) & 0.23 & (1,211) & 1.00 & (283.7) & 0.15 & (844.5) & 0.09 & (1,409) & 1.00 & (123.5) \\
sanr400-0.5 & 0.28 & (9.9) & 0.31 & (8.8) & 1.00 & (2.7) & 0.16 & (9.2) & 0.12 & (12.7) & 1.00 & (1.5) \\ 
sanr400-0.7 & 0.10 & (7,292) & 0.28 & (2,544) & 1.00 & (711.9) & 0.14 & (2,698) & 0.10 & (3,737) & 1.00 & (365.6) \\ \hline
ratio (total)    & 0.12 & (9,033) & 0.26 & (4,202) & 1.00 & (1,098) & 0.14 & (3,937) & 0.10 & (5,622) & 1.00 & (539.4) \\ \hline
\end{tabular}
\end{scriptsize}
\end{center}
\caption{Calibration experiments using DIMACS instances, two algorithms (MCSa and BBMC) and three machines. For each algorithm
we tabulate for each machine (Fais, Daleview, Cyprus) the ratio of run time against the reference machine (Cyprus) and
(in brackets) the actual run time in seconds on that machine.}
\label{calibration}
\end{table}

Table \ref{calibration} shows the results of the calibration experiments. Tabulated are a subset of DIMACS instances that
took more than 1 second and less than 2 hours to solve using MCSa on our second slowest machine (Fais). Run times are tabulated in
seconds and the actual ratio of Cyprus-time over Fais-time (expected to be 0.41) is given as well as Cyprus-time over
Daleview-time (expected to be 0.50) for each data point. Two algorithms are used, MCSa and BBMC. The last row of
Table \ref{calibration} gives the 
relative performance ratios computed using the sum of the run times in the table. Referring back to Table \ref{rosettaStone} we expect a 
Cyprus/Fais ratio of 0.41 but empirically get
0.12 when using MCSa and 0.14 when using BBMC. We expect a Cyprus/Daleview ratio of 0.50 but empirically
get an average 0.26 with MCSa and 0.10 with BBMC. 

The conversion factors in Table \ref{rosettaStone} consistently
over-estimate the speed of Fais and Daleview. For example, we would expect MCSa applied to brock200-1 on Fais to have a run time
of $19.343 \times 0.41 \approx 7.9$ seconds on Cyprus. In fact it takes 4.8 seconds. If we use the derived ratio
in the last row of Table \ref{calibration} we get $19.343 \times 0.12 \approx 2.3$ seconds. 
As another example consider san1000 using BBMC on Daleview. We would expect this to take $54.816 \times 0.50 \approx 27.4$
seconds on Cyprus. In fact it takes 5.9 seconds! If we use the conversion ratio from the last row of Table \ref{calibration}
we get a more accurate estimate $54.816 \times 0.10 \approx 5.5$ seconds.

\begin{table}
\begin{center}
\begin{scriptsize}
\begin{tabular}{|l|c c|c c|c c|c c|c c|c c|} \hline 
\multicolumn{1}{|c|} {} & \multicolumn{6}{|c|}{Cliquer} & \multicolumn{6}{|c|}{dfmax}\\
\multicolumn{1}{|c|} {instance} & \multicolumn{2}{|c}{Fais} & \multicolumn{2}{c}{Daleview} & \multicolumn{2}{c|}{Cyprus} & \multicolumn{2}{|c}{Fais} & \multicolumn{2}{c}{Daleview} & \multicolumn{2}{c|}{Cyprus} \\ \hline
brock200-1 & 0.66 & (9.8) & 0.43 & (18.7) & 1.00 & (6.5) & 0.39 & (25.2) & 0.42 & (23.0) & 1.00 & (9.7) \\
brock200-4 & 0.64 & (0.7) & 0.47 & (1.2) & 1.00 & (0.4) & 0.41 & (1.5) & 0.46 & (1.4) & 1.00 & (0.6) \\
p-hat1000-1 &0.62 & (1.8) & 0.36 & (3.0) & 1.00 & (1.1) & 0.41 & (1.7) & 0.45 & (1.5) & 1.00 & (0.7) \\
p-hat700-1 & 0.67 & (0.2) & 0.37 & (0.3) & 1.00 & (0.1) & --- & --- & --- & --- & --- & --- \\
san1000 & 0.75 & (0.1) & 0.30 & (0.3) & 1.00 & (0.1) & --- & --- & --- & --- & --- & --- \\
san200-0.7-1 & 0.48 & (1.8) & 0.20 & (4.2) & 1.00 & (0.8) & --- & --- & --- & --- & --- & --- \\
san200-0.9-2 & 0.61 & (18.9) & 0.21 & (54.0) & 1.00 & (11.3) & --- & --- & --- & --- & --- & --- \\
san400-0.7-3 & 0.62 & (6.8) & 0.26 & (16.1) & 1.00 & (4.2) & --- & --- & --- & --- & --- & --- \\
sanr200-0.7 & 0.65 & (2.9) & 0.36 & (5.3) & 1.00 & (1.9) & 0.40 & (5.2) & 0.44 & (4.8) & 1.00 & (2.1) \\
sanr400-0.5 & 0.62 & (1.5) & 0.38 & (2.4) & 1.00 & (0.9) & 0.41 & (3.6) & 0.47 & (3.1) & 1.00 & (1.5) \\ \hline
ratio (total) & 0.62 & (44.3) & 0.26 & (105.5)& 1.00 & (27.6) & 0.39 & (37.1) & 0.43 & (33.8) & 1.00 & (14.6) \\ \hline
\end{tabular}
\end{scriptsize}
\end{center}
\caption{Calibration experiments for Cliquer and dfmax.}
\label{cliquerCalibration}
\end{table}


But maybe this is because we have used a C program (dfmax) to calibrate a Java program. Would we get a reliable calibration if a C 
program was used? \"{O}sterg\aa{}rd's Cliquer program was downloaded and compiled on our three machines and run against
DIMACS benchmarks, i.e. the experiments in Table \ref{calibration} were repeated using Cliquer and dfmax with a different, and easier, 
set of problems (as these algorithms are much slower). The results are shown
in Table \ref{cliquerCalibration}\footnote{An entry --- was a run of dfmax that was terminated after 2 minutes.}. What we see is an actual 
scaling factor of 0.62 for Cliquer on Fais when dfmax predicts 0.41 
and for Cliquer on Daleview 0.26 when we expect 0.50; again we see that the rescaling procedure fails. The last three
columns show a dfmax calibration using problems other than the r* benchmarks and here we see an
error of about 5\% on Fais (expected 0.41, actual 0.39) and  about 16\% on Daleview (expected 0.50, actual 0.43). 
Therefore it appears that rescaling results using dfmax and the five r* benchmarks is not a safe procedure and can result in wrong
conclusions being drawn regarding the relative performance of algorithms.

\section{Relative algorithmic performance on different machines}
\vspace{-1.5mm}
But is it even safe to draw conclusions on our algorithms when we base those conclusions on experiments
performed on a single machine? Previously, in Table~\ref{tableMCSvBBMC} we compared MCSa against BBMC on our
reference machine Cyprus and concluded that BBMC was typically twice as fast as MCSa. Will that hold on
Fais and on Daleview? 

\begin{table}
\begin{center}
\begin{scriptsize}
\begin{tabular}{|l|c|c|c|}  \hline
instance & ~~~Fais~~~ & Daleview & Cyprus \\ \hline
brock200-1 & 1.26  & 0.70  & 2.03  \\ 
brock200-4 & 1.17  & 0.72  & 2.35  \\ 
hamming10-2 & 3.10  & 3.24  & 2.21  \\ 
hamming8-4 & 1.16  & 0.86  & 1.24  \\ 
johnson16-2-4 & 1.23  & 0.61  & 1.66  \\ 
MANN-a27 & 2.62  & 1.93  & 4.12  \\ 
p-hat1000-1 & 1.01  & 0.79  & 1.80  \\ 
p-hat1500-1 & 0.86  & 0.72  & 1.52  \\ 
p-hat300-3 & 1.28  & 0.85  & 2.36  \\ 
p-hat500-2 & 1.32  & 0.90  & 2.56  \\ 
p-hat700-1 & 1.38  & 0.88  & 1.86  \\ 
san1000 & 1.63  & 1.63  & 1.76  \\ 
san200-0.9-1 & 1.17  & 0.71  & 1.93  \\ 
san200-0.9-2 & 1.61  & 0.88  & 2.36  \\ 
san400-0.7-1 & 1.23  & 0.94  & 2.04  \\ 
san400-0.7-2 & 1.18  & 0.84  & 2.12  \\ 
san400-0.7-3 & 1.14  & 0.87  & 2.10  \\ 
sanr200-0.7 & 1.24  & 0.70  & 1.95  \\ 
sanr200-0.9 & 1.48  & 0.86 & 2.30 \\
sanr400-0.5 & 1.08  & 0.69 & 1.85  \\ 
sanr400-0.7 & 2.70 & 0.68 & 1.95 \\ \hline
\end{tabular}
\end{scriptsize}
\end{center}
\caption{Does hardware affect relative algorithmic performance? 
Values greater than 1 imply BBMC is faster than MCSa, less than 1 MCSa is faster.}
\label{calibrationAlg}
\end{table}

Table \ref{calibrationAlg} takes the data from Table \ref{calibration} and divides the
run time of MCSa by BBMC for each instance on our three machines. 
On Fais BBMC is rarely more than 50\% faster than MCSa and on Daleview BBMC is slower than MCSa more often than not! 
If experiments were performed
only on Daleview using only the DIMACS instances we might draw entirely different conclusions and claim that BBMC is slower 
than MCSa!\footnote{The -server and -client options were also tried. The -server option sometimes
gave speedups of a factor of 2 sometimes a factor of 0.5, and this can also affect relative algorithmic performance.}

\section{Changing language}
What happens if we change the implementation language? Our two algorithms were recoded in C++ and the experiments in Table 
\ref{calibration} repeated. On Fais and Daleview our code was compiled with g++ 3.4 and g++ 4.7 on our reference machine Cyprus.
Again, we see the standard rescaling procedure overestimating the speeds of Fais and Daleview. We also see again that
the rescaling ratio varies depending on the algorithm used: the Cyprus/Daleview ratio is 0.29 for MCSa but
0.14 for BBMC (the same phenomena we see in Table \ref{calibration}).

\begin{table}
\begin{center}
\begin{scriptsize}
\begin{tabular}{|l|c c|c c|c c|c c|c c|c c|} \hline 
\multicolumn{1}{|c|} {} & \multicolumn{6}{|c|}{MCSa} & \multicolumn{6}{|c|}{BBMC}\\
\multicolumn{1}{|c|} {instance} & \multicolumn{2}{|c}{Fais} & \multicolumn{2}{c}{Daleview} & \multicolumn{2}{c|}{Cyprus} & \multicolumn{2}{|c}{Fais} & \multicolumn{2}{c}{Daleview} & \multicolumn{2}{c|}{Cyprus} \\ \hline
brock200-1 & 0.17 & (13.0) & 0.03 & (6.9) & 1.00 & (2.2) & 0.13 & (2.8) & 0.14 & (2.6) & 1.00 & (0.37) \\ 
brock200-4 & 0.58 & (1.4) & 0.35 & (0.66) & 1.00 & (0.23) & 0.14 & (0.29) & 0.15 & (0.26) & 1.00 & (0.04) \\ 
hamming10-2 & 0.17 & (0.77) & 0.10 & (1.3) & 1.00 & (0.13) & 0.23 & (0.13) & 3.0 & (0.01) & 1.00 & (0.03) \\ 
hamming8-4 & 0.19 & (1.2) & 0.34 & (0.68) & 1.00 & (0.23) & 0.13 & (0.31) & 0.13 & (0.30) & 1.00 & (0.04) \\ 
johnson16-2-4 & 0.08 & (3.1) & 0.27 & (0.97) & 1.00 & (0.26) & 0.10 & (0.49) & 0.12 & (0.42) & 1.00 & (0.05) \\ 
MANN-a27 & 0.28 & (10.2) & 0.20 & (14.5) & 1.00 & (2.9) & 0.15 & (1.7) & 0.14 & (1.8) & 1.00 & (0.25) \\ 
p-hat1000-1 & 0.06 & (15.4) & 0.29 & (3.4) & 1.00 & (0.99) & 0.15 & (1.6) & 0.14 & (1.7) & 1.00 & (0.24) \\ 
p-hat1500-1 & 0.06 & (153.2) & 0.27 & (36.1) & 1.00 & (9.8) & 0.18 & (16.2) & 0.15 & (19.4) & 1.00 & (2.9) \\ 
p-hat300-3 & 0.20 & (27.7) & 0.27 & (20.4) & 1.00 & (5.5) & 0.14 & (7.4) & 0.14 & (7.3) & 1.00 & (1.0) \\ 
p-hat500-2 & 0.19 & (7.5) & 0.29 & (4.8) & 1.00 & (1.4) & 0.14 & (1.7) & 0.15 & (1.6) & 1.00 & (0.24) \\ 
p-hat700-1 & 0.08 & (1.7) & 0.30 & (0.47) & 1.00 & (0.14) & 0.16 & (0.25) & 0.14 & (0.28) & 1.00 & (0.04) \\ 
san1000 & 0.20 & (27.5) & 0.22 & (25.5) & 1.00 & (5.5) & 0.17 & (10.6) & 0.17 & (10.4) & 1.00 & (1.8) \\ 
san200-0.9-1 & 0.20 & (2.6) & 0.29 & (1.8) & 1.00 & (0.52) & 0.14 & (0.66) & 0.15 & (0.60) & 1.00 & (0.09) \\ 
san200-0.9-2 & 0.25 & (9.3) & 0.27 & (8.4) & 1.00 & (2.3) & 0.15 & (2.3) & 0.16 & (2.1) & 1.00 & (0.34) \\ 
san400-0.7-1 & 0.16 & (6.1) & 0.28 & (3.6) & 1.00 & (0.99) & 0.13 & (1.6) & 0.14 & (1.5) & 1.00 & (0.21) \\ 
san400-0.7-2 & 0.19 & (47.6) & 0.31 & (28.9) & 1.00 & (8.9) & 0.14 & (13.2) & 0.15 & (12.7) & 1.00 & (1.9) \\ 
san400-0.7-3 & 0.19 & (27.7) & 0.33 & (16.3) & 1.00 & (5.4) & 0.15 & (8.2) & 0.16 & (7.6) & 1.00 & (1.2) \\ 
sanr200-0.7 & 0.17 & (3.7) & 0.35 & (1.8) & 1.00 & (0.63) & 0.13 & (0.78) & 0.14 & (0.72) & 1.00 & (0.10) \\ 
sanr200-0.9 & 0.22 & (581.3) & 0.35 & (498.3) & 1.00 & (129.1) & 0.14 & (133.5) & 0.16 & (119.9) & 1.00 & (18.6) \\
sanr400-0.5 & 0.10 & (12.4) & 0.35 & (3.7) & 1.00 & (1.3) & 0.13 & (2.0) & 0.13 & (1.9) & 1.00 & (0.25) \\ 
sanr400-0.7 & 0.13 & (2,706.7) & 0.31 & (1094.8) & 1.00 & (343.2) & 0.13 & (537.8) & 0.13 & (527.3) & 1.00 & (68.6) \\ \hline
ratio (total)    & 0.14 & (3,660.1) & 0.29 & (1,773.3) & 1.00 & (521.6) & 0.13 & (743.5) & 0.14 & (720.4) & 1.00 & (98.3) \\ \hline
\end{tabular}
\end{scriptsize}
\end{center}
\caption{Calibration experiments in C++.}
\label{cCalibration}
\end{table}

But what of the relative performance of MCSa with respect to BBMC? In Tables \ref{calibrationAlg} we saw that on Daleview BBMC 
was slower than MCSa. In our C++ encoding this does not occur and on all of our machines BBMC is faster than MCSa.

\section{Conclusion}
\vspace{-1.5mm}
In this day and age, the day of the download, we expect that code used in experiments will be available on the web
and that we can download it, or that we can email authors and personally ask for their code. In the experiments here code was available 
from Patric R. J. \"{O}sterg\aa{}rd (Cliquer), Janez Konc and Du\u{s}anka Jane\u{z}i\u{c} (MaxCliqueDyn), and of course dfmax.
Jean-Charles R\'{e}gin's algorithm is implemented in ILOG Solver, a commercial toolkit, so this was excluded from this study. 
No source code or programs were available for BB-MaxClique or for MCQ, MCR and MCS. Therefore to include
these algorithms in a study two options are available: roll your own or rescale. Here, we rolled our own\footnote{... as did Renato Carmo and
Alexandre P. Z{\"u}ge \cite{carmoZuge}, although they omitted BBMC.}.
But if we are put in the unfortunate position where we are forced to rescale, what are the consequences?
Rescaling is unsafe and should not be trusted.  Nevertheless, this procedure is still in use today \cite{MCMD14}.

There are, unfortunately, examples in the literature of how not to do it. A closer inspection of Figure 1 shows
that some authors have gone so far as to rescale their own results. Furthermore, a close reading of the literature
shows cases where small samples (size 10) of random problems are used as benchmarks. 
Authors then compare rescaled results across publications that use a different small set of random problems.
There are also numerous examples of cherry picking where some DIMACS instances (66 in all) are omitted from studies with no explanation.

In our study of MCSa and BBMC we have used three machines and two implementation languages. We had the following 
choices available to us: choose a machine for each algorithm, either the same (3 ways) or different (3 ways and use rescaling);
implement BBMC in C++ or Java (2 ways) and the same for MCSa (2 ways). This gives us $(3+3) \times 2 \times 2$ possible 
designs for our experiments. We have explored only a subset of these 24 designs, but the data presented in our Tables allows the reader 
to make what he will. In a nutshell, from our study we can draw whatever conclusion we like.

The Java code used in this study is available online at \url{http://www.dcs.gla.ac.uk/~pat/maxClique} along with instructions on how to 
run the code, the DIMACS instances, random problem generator and run time results. 
The C++ code used in this study is available at \url{https://github.com/ciaranm/multithreadedmaximumclique/archive/master.zip}
\vspace{-2mm}

\section*{Acknowledgements}
Ciaran McCreesh encoded the algorithms in C++ and helped with the writing of this report. 



\begin{thebibliography}{10}

\bibitem{DIMACS}
{DIMACS}.
\newblock ftp://dimacs.rutgers.edu/pub/challenge/graph/benchmarks/clique.

\bibitem{carmoZuge}
Renato Carmo and Alexandre~P. Z{\"u}ge.
\newblock Branch and bound algorithms for the maximum clique problem under a
  unified framework.
\newblock {\em J. Braz. Comp. Soc.}, 18(2):137--151, 2012.

\bibitem{carraghanPardalos90}
Randy Carraghan and Panos~M. Pardalos.
\newblock {An exact algorithm for the maximum clique problem}.
\newblock {\em Operations Research Letters}, 9:375--382, 1990.

\bibitem{MCMD14}
Ricardo~C. Corr{\^{e}}a, Philippe Michelon, Bertrand~Le Cun, Thierry Mautor,
  and Diego~Delle Donne.
\newblock A bit-parallel russian dolls search for a maximum cardinality clique
  in a graph.
\newblock {\em CoRR}, abs/1407.1209, 2014.

\bibitem{fahle}
Torsten Fahle.
\newblock {Simple and Fast: Improving a Branch-and-Bound Algorithm for Maximum
  Clique}.
\newblock In {\em Proceedings {ESA 2002}, LNCS 2461}, pages 485--498, 2002.

\bibitem{gareyJohnson}
M.R. Garey and D.S. Johnson.
\newblock {\em Computers and Intractability}.
\newblock W.H. Freeman and Co, 1979.

\bibitem{dimacs93}
David~J. Johnson and Michael~A. Trick, editors.
\newblock {\em Cliques, Coloring, and Satisfiability: Second DIMACS
  Implementation Challenge, Workshop, October 11-13, 1993}.
\newblock American Mathematical Society, Boston, MA, USA, 1996.

\bibitem{howTo}
David~S. Johnson.
\newblock A theoretician's guide to the experimental analysis of algorithms,
  1996.

\bibitem{Konc_Janezic_2007}
Janez Konc and Du\u{s}anka Jane\u{z}i\u{c}.
\newblock An improved branch and bound algorithm for the maximum clique
  problem.
\newblock {\em MATCH Communications in Mathematical and Computer Chemistry},
  58:569--590, 2007.

\bibitem{aaai2010}
Chu~Min Li and Zhe Quan.
\newblock An efficient branch-and-bound algorithm based on maxsat for the
  maximum clique problem.
\newblock In {\em AAAI'10}, pages 128--133, 2010.

\bibitem{prjo2002}
Patric R.~J. \"{O}sterg\aa{}rd.
\newblock A fast algorithm for the maximum clique problem.
\newblock {\em Discrete Applied Mathematics}, 120:197--207, 2002.

\bibitem{pardalosRodgers92}
Panos~M. Pardalos and Gregory~P. Rodgers.
\newblock {A Branch and Bound Algorithm for the Maximum Clique Problem}.
\newblock {\em Computers and Operations Research}, 19:363--375, 1992.

\bibitem{exactProsser}
Patrick Prosser.
\newblock Exact algorithms for maximum clique: A computational study.
\newblock {\em Algorithms}, 5(4):545--587, 2012.

\bibitem{regin2003}
Jean-Charles R\'{e}gin.
\newblock {Using Constraint Programming to Solve the Maximum Clique Problem}.
\newblock In {\em Proceedings {CP 2003}, LNCS 2833}, pages 634--648, 2003.

\bibitem{segundo2011b}
Pablo~San Segundo, Fernando Matia, Diego Rodr\'{i}guez-Losada, and Miguel
  Hernando.
\newblock {An improved bit parallel exact maximum clique algorithm}.
\newblock {\em Optimization Letters}, 2011.

\bibitem{segundo2011}
Pablo~San Segundo, Diego Rodr\'{i}guez-Losada, and August\'{i}n Jim\'{e}nez.
\newblock {An exact bit-parallel algorithm for the maximum clique problem}.
\newblock {\em Computers and Operations Research}, 38:571--581, 2011.

\bibitem{sewell98}
E.~C. Sewell.
\newblock A branch and bound algorithm for the stability number of a sparse
  graph.
\newblock {\em INFORMS Journal on Computing}, 10(4):438--447, 1998.

\bibitem{tomita2007}
E.~Tomita and Toshikatsu Kameda.
\newblock An efficient branch-and-bound algorithm for finding a maximum clique
  and computational experiments.
\newblock {\em Journal of Global Optimization}, 37:95--111, 2007.

\bibitem{tomita2003}
E.~Tomita, Y.~Sutani, T.~Higashi, S.~Takahashi, and M.~Wakatsuki.
\newblock An efficient branch-and-bound algorithm for finding a maximum clique.
\newblock In {\em DMTC 2003, LNCS 2731}, pages 278--289, 2003.

\bibitem{tomita2010}
E.~Tomita, Y.~Sutani, T.~Higashi, S.~Takahashi, and M.~Wakatsuki.
\newblock A simple and faster branch-and-bound algorithm for finding maximum
  clique.
\newblock In {\em WALCOM 2010, LNCS 5942}, pages 191--203, 2010.

\bibitem{welshPowell}
D.J.A. Welsh and M.B. Powell.
\newblock {An upper bound for the chromatic number of a graph and its
  application to timetabling problems}.
\newblock {\em The Computer Journal}, 10(1):85--86, 1967.

\bibitem{wood97}
David~R. Wood.
\newblock {An algorithm for finding a maximum clique in a graph}.
\newblock {\em Operations Research Letters}, 21:211--217, 1997.

\end{thebibliography}

\end{document}